\documentclass[a4paper, 11pt]{article}

\usepackage{graphicx}
\usepackage{epsf,amsmath,bbold,amsfonts,stmaryrd}
\usepackage{multirow}
\usepackage{makecell} 
\usepackage[utf8]{inputenc}
\usepackage{mathrsfs}
\usepackage{appendix}
\usepackage{amssymb}
\usepackage{float}
\usepackage{color}
\usepackage{cite}
\usepackage{hyperref}
\hypersetup{pageanchor=false}
\usepackage{indentfirst}
\usepackage{url}
\usepackage{float}
\usepackage{caption}
\usepackage[numbers,square,comma,sort&compress,merge]{natbib}

\usepackage{subfigure}

\usepackage{ulem}

\def\be{\begin{equation}}
\def\ee{\end{equation}}

\def\bea{\begin{eqnarray}}
\def\eea{\end{eqnarray}}

\def\ba{\begin{array}}
\def\ea{\end{array}}

\def\bc{\begin{center}}
\def\ec{\end{center}}

\def\bl{\begin{flushleft}}
\def\el{\end{flushleft}}

\def\br{\begin{flushright}}
\def\er{\end{flushright}}

\def\bi{\begin{itemize}}
\def\ei{\end{itemize}}

\def\bt{\begin{tabular}}
\def\et{\end{tabular}}

\numberwithin{equation}{section}

\begin{document}

\title{Broadened Lensing Rings of Compact Boson Stars: Enhanced Imprint of Accretion Flow in Images and Visibilities}

\author{Xiangyu Wang$^{1}$, Xinyu Wang$^{2,3}$, Minyong Guo$^{2, 3\ast}$, Hai-Qing Zhang$^{1, 4\dagger}$
}
\date{}

\maketitle

\vspace{-10mm}

\begin{center}
{\it
	$^1$ Center for Gravitational Physics, Department of Space Science, Beihang University, Beijing
	100191, China\\\vspace{4mm}
	$^2$Department of Physics, Beijing Normal University, Beijing 100875, China\\\vspace{4mm}

	$^3$Key Laboratory of Multiscale Spin Physics (Beijing Normal University), Ministry of Education, Beijing 100875, China\\\vspace{4mm}
	$^4$Peng Huanwu Collaborative Center for Research and Education, Beihang University, Beijing
	100191, China\\\vspace{4mm}
}
\end{center}

\vspace{8mm}

\begin{abstract}

In this work, we systematically study the gravitational lensing properties and observational signatures of compact boson stars. Unlike black holes, the photon effective potential of a compact boson star develops a nearly flat region, whose width increases with the compactness of the star. This flat structure significantly broadens the range of impact parameters that can produce large-angle deflections, leading to noticeably wider lensing rings of all orders. Photons constituting these rings traverse more complex paths, rendering the resulting images more sensitive to the spatial distribution of the accretion flow. Ray tracing results show that, compared to black hole models, the image topology and visibility amplitudes of compact boson stars exhibit a stronger dependence on the accretion flow structure. These results highlight qualitative differences in the observational properties of compact boson stars and black holes.

\end{abstract}

\vfill{
	\footnotesize $\ast$ Corresponding author: minyongguo@bnu.edu.cn
	
	\footnotesize $\dagger$ Corresponding author: hqzhang@buaa.edu.cn

}

\maketitle

\newpage
\baselineskip 18pt

\section{Introduction}
\label{sec1}

Event Horizon Telescope (EHT) observations of M87$^*$ and Sgr~A$^*$ at 1.3\,mm wavelength have for the first time resolved the bright emission ring surrounding the central brightness depression, whose morphology is in striking agreement with the predictions for the light deflection from general 
relativty in strong gravitational fields\cite{Gillessen:2008qv,Genzel:2010zy,GRAVITY:2018ofz,GRAVITY:2020gka,Do:2019txf,EventHorizonTelescope:2022wkp,Yan:2022fkr,Zhang:2024fpm,EventHorizonTelescope:2019ggy,EventHorizonTelescope:2019pgp,EventHorizonTelescope:2019ths,EventHorizonTelescope:2019dse,EventHorizonTelescope:2022wkp}.
The observations show that although the overall angular scale of the ring has been successfully resolved, its width and finer substructure remain close to the limit of the current angular resolution\cite{EventHorizonTelescope:2022wkp,EventHorizonTelescope:2019dse}.
From a theoretical perspective, this bright ring is not a single emission structure but rather a nested sequence of photon rings formed by photon orbits containing different numbers of half-orbits around the central object\cite{Johannsen:2015mdd,Gralla:2019xty}.
These higher-order photon rings are exponentially sensitive to the spacetime geometry\cite{Johnson:2019ljv} and are therefore regarded as important observational windows for testing strong-field general relativity and probing the nature of compact objects.

Although EHT observations overall support the existence of supermassive black holes, with the current resolution and imaging uncertainties\cite{Psaltis:2014mca}, it remains impossible to completely exclude the possibility that the central objects are horizonless compact alternatives\cite{Cunha:2018acu}.
Indeed, previous studies have shown that, within certain parameter ranges, exotic compact objects can also produce ring-like imaging features that closely resemble black hole shadows \cite{Vincent:2015xta,Rosa:2022tfv,Herdeiro:2016gxs}.
From a theoretical standpoint, this problem is of great importance: the spacetime singularity inside a black hole signifies the breakdown of classical general relativity\cite{Penrose:1964wq}, while the information-loss puzzle associated with the event horizon remains unresolved\cite{Hawking:1976ra,Preskill:1992tc}.
Therefore, constructing and investigating compact object models that are free of singularities and event horizons not only helps to understand spacetime structures in the strong-gravity limit\cite{Morris:1988cz,Mazur:2001fv,Mathur:2005zp} but may also provide observational windows into quantum gravity effects\cite{Giddings:2013jra,Dvali:2011aa}.

Among many alternative models, boson stars are one of the most representative proposals.
These objects are composed of self-gravitating scalar fields, everywhere regular at the classical level, and do not possess an event horizon\cite{Schunck:2003kk,Cardoso:2019rvt}.
In recent years, extensive work has been devoted to the stability\cite{Gleiser:1988rq}, dynamical evolution\cite{Seidel:1990jh,Liebling:2012fv}, lensing structure\cite{herdeiro2021imitation}, GRMHD simulation\cite{Torres:2002td,Guzman:2005bs,Guzman:2009zz,Macedo:2013jja,Palenzuela:2017kcg,Olivares:2018abq,Olivares-Sanchez:2024dfh,Gjorgjieski:2024csb} and accretion flow imaging of boson stars\cite{Yang:2025usj,Zeng:2025fjg,Zeng:2025nmu, Rosa:2022toh,Rosa:2023qcv,Rosa:2024eva,Aimar:2025uia,Huang:2024gtu}.
Among these, studies on gravitational lensing and accretion flow imaging indicate that the imaging characteristics of boson stars are closely related to their compactness. In the low-compactness regime, higher-order lensing rings have not yet formed and the image structure is relatively simple, which is  dominated by photons directly emitted from the radiation region. However, as the compactness increases, the strong gravitational lensing effect gradually intensifies, and photon orbits that undergo multiple windings begin to contribute more radiation, causing richer lensing features to emerge in the image. These structural changes may further manifest themselves in the imaging morphology and interferometric visibility, thereby giving rise to the potential observable differences. At present, however, systematic studies of this process remain relatively scarce.

Based on the above background, this paper focuses on the gravitational lensing and imaging properties of compact solitonic boson stars in the strong-coupling parameter regime.
We focus on the structure of the photon effective potential and its influence on orbital dynamics, and find that in compact configurations the effective potential develops a pronounced flat region in the inner region, which suppresses the radial escape of photons and prolongs their residence time in the strong-field region.
This mechanism leads to a systematic broadening of all-order lensing rings, making the resulting image morphology more sensitive to the spatial structure of the accretion flow.
On this basis, we further analyze the interferometric visibility amplitudes and find that, compared with the black hole case, compact boson stars exhibit a remarkably enhanced sensitivity to accretion flow models.

The paper is organized as follows. Sec.~\ref{sec2} presents the numerical framework, including the construction of boson star spacetimes, the accretion flow radiation model, and the method for computing visibility amplitudes. Sec.~\ref{sec3} analyzes the geodesic structure of boson stars, focusing on the influence of the flat effective potential on photon trajectories and deflection angles. Sec.~\ref{sec4} presents the results of imaging and visibility simulations. Finally, the main conclusions are summarized in Sec.~\ref{sec5}. 

\section{Basic Setup}\label{sec2}
To investigate the imaging features of boson stars, a consistent treatment of the underlying spacetime geometry, radiative transfer, and interferometric observables is required. In this section, we present the numerical framework adopted throughout this work. We first briefly review the boson star spacetime and describe the configurations used in our study. We then introduce the thick-disk emission model employed in the ray-tracing calculations, followed by the procedure of constructing visibility amplitudes from the simulated intensity maps.

The framework developed here serves as the basis for all subsequent analyses, including the study of lensing structures, image morphology, and Fourier domain signatures.
\subsection{Boson Star Spacetime}
We first introduce the boson star configurations considered in this work. The coupled scalar–gravitational system is governed by the action
\begin{equation}
	\mathcal{S}=\int d^4x\sqrt{-g}
	\left[
	\frac{R}{2\kappa}
	-\frac{1}{2}\nabla_a\Phi^*\nabla^a\Phi
	-V(|\Phi|^2)
	\right],
\end{equation}
where $R$ is the Ricci scalar, $\kappa=8\pi G$, and $\Phi$ denotes the complex scalar field. Throughout this paper, we adopt the natural units with $G=c=1$. The soliton scalar self-interaction potential is 
\begin{equation}
	V(|\Phi|^2)
	=
	\mu^2|\Phi|^2
	\left(
	1+\frac{|\Phi|^2}{\alpha^2}
	\right)^2,
\end{equation}
where $\mu$ is the boson mass and $\alpha$ characterizes the self-interaction strength.

Varying the action with respect to the metric and scalar field yields the Einstein and Klein-Gordon equations, respectively
\bea
\begin{aligned}
	&R_{ab}-\frac{1}{2}g_{ab}R
	=
	\kappa T_{ab},
	\\
	&\nabla_a\nabla^a\Phi
	=
	\Phi\frac{dV}{d|\Phi|^2},
\end{aligned}
\label{gphieoms}
\eea
where the energy-momentum tensor is given by
\begin{equation}
	T_{ab}
	=
	\nabla_a\Phi^*\nabla_b\Phi
	+
	\nabla_b\Phi^*\nabla_a\Phi
	-
	g_{ab}
	\left(
	\nabla_c\Phi^*\nabla^c\Phi+V
	\right).
\end{equation}

Assuming a static and spherically symmetric spacetime, we adopt the metric and scalar-field ansatz
\begin{equation}
	ds^2
	=
	-A(r)dt^2
	+\frac{dr^2}{B(r)}
	+r^2d\Omega^2,
	\label{dsansatz}
\end{equation}
and
\begin{equation}
	\Phi(r,t)=\psi(r)e^{-i\omega t},
	\label{phiansatz}
\end{equation}
where $\omega$ is the scalar-field frequency. Substituting Eqs.~\eqref{dsansatz} and \eqref{phiansatz} into Eq.~\eqref{gphieoms}, we obtain
\bea
\begin{aligned}
	A'
	&=
	\frac{A(1-B)}{Br}
	+
	\kappa r
	\left(
	\frac{\omega^2\psi^2}{B}
	+
	A(\psi')^2
	-
	\frac{A}{B}V
	\right),
	\\
	B'
	&=
	\frac{1-B}{r}
	-
	\kappa r
	\left(
	\frac{\omega^2\psi^2}{A}
	+
	B(\psi')^2
	+
	V
	\right),
	\\
	\psi''
	&+
	\left[
	\frac{2}{r}
	+
	\frac{1}{2}
	\left(
	\frac{A'}{A}
	+
	\frac{B'}{B}
	\right)
	\right]\psi'
	+
	\frac{1}{B}
	\left(
	\frac{\omega^2}{A^2}
	-
	\frac{dV}{d\psi}
	\right)\psi
	=
	0.
\end{aligned}
\label{BSeom}
\eea

Regularity at the origin requires
\begin{equation}
	A(0)=A_0,\qquad
	B(0)=1,\qquad
	\psi(0)=\psi_0,
	\label{bound}
\end{equation}
while asymptotic flatness imposes
\begin{equation}
	A(r),\,B(r)\xrightarrow[r\to\infty]{}1-\frac{2M}{r},
	\qquad
	\psi(r)\xrightarrow[r\to\infty]{}0.
\end{equation}

The system is solved numerically using the shooting method. Owing to the rescaling freedom of the time coordinate, $A_0$ can be fixed initially and later rescaled to satisfy the asymptotic boundary condition. The boson star solutions are therefore mainly characterized by the parameters $(\alpha,\psi_0)$.
For given $(\alpha, \psi_0)$, the scalar-field frequency $\omega$ is determined as an eigenvalue by requiring the scalar field to decay at spatial infinity. The ADM mass $M$ is extracted from the asymptotic behavior. Since boson stars do not possess a sharply defined surface, we define an effective radius $R$ by the radial coordinate where the enclosed mass reaches $98\%$ of the total ADM mass. Finally, the compactness, defined as the ratio of the total mass to the effective radius $M/R$, quantifies how compact the boson star is.
\begin{table}[!htbp]
	\centering
	\renewcommand{\arraystretch}{1.3}
	\begin{tabular}{l|c|c|c|c|c|c}
		\hline\hline
		Model & $\alpha$ &$\psi_0$ & $\mu M$ & $\mu R$ & $\omega/\mu$ & $R/M$ \\
		\hline 
		BS1 & 0.08 & 0.0912 & 1.06 & 4.92 & 0.353 & 4.67 \\ 
		\hline 
		BS2 & 0.08 & 0.0938 & 1.18 & 4.85 & 0.323 & 4.11 \\ 
		\hline
		BS3 & 0.08 & 0.0973 & 1.16 & 4.63 & 0.320 & 3.99  \\ 
		\hline
		BS4 & 0.08 & 0.1008 & 1.11 & 4.42 & 0.324 & 3.96 \\ 
		\hline 
		BS5 & 0.06 & 0.0708 & 2.16 & 7.28 & 0.226 & 3.36 \\ 
		
		\hline\hline
	\end{tabular}
	\caption{Parameters for different boson star model.}
	\label{para}
\end{table}
	This paper focuses on the imaging features of compact boson stars. Five representative configurations are selected for investigation, with their characteristic parameters and labels listed in Tab.~\ref{para}. These configurations, denoted BS1 through BS5, are arranged in order of increasing compactness. Among of these, the most compact one, BS5, is sufficiently 
	compact to devlop a photon ring. 
\subsection{Accretion Flow and Radiative Transfer Model}

To model the electromagnetic emission surrounding the boson star, we adopt a radiatively inefficient accretion flow (RIAF) model. The accretion flow is assumed to be stationary, axisymmetric, geometrically thick, and optically thin. The geometrically thick disk configuration is identical to that employed in \cite{Wang:2025qpv}. In this work, we neglect radial infall motion and consider only rotational motion of the emitting plasma. The fluid four-velocity is therefore taken as
\begin{equation}
	u^\mu = u^t (1,\,0,\,0,\,\Omega), 
\end{equation}
 We employ cylindrical coordinates $(R,\,z)$ with $R = r\sin\theta$ the cylindrical radius and $z = r\cos\theta$ the height above the equatorial plane. The angular velocity $\Omega = u^\phi / u^t$ is specified by
\begin{equation}
	\Omega = -\frac{g_{tt}}{g_{\phi\phi}}\cdot\frac{l_0 R^{3/2}}{C + R}.
\end{equation}
 In numerics, the model parameters are set to $l_0 = C = 1$, which corresponds to a purely rotational, idealised Keplerian-type configuration. 
 
 Unlike black holes, the angular velocity of timelike circular orbits around a boson star can attain a maximum at a finite radius.
 We denote this radius by $r_\Omega$, defined by the condition $\mathrm{d}\Omega/\mathrm{d}r = 0$.
 Consequently, the rotation profile satisfies $\mathrm{d}\Omega/\mathrm{d}r < 0$ outside $r_\Omega$, while the opposite holds in the inner region.
 As pointed out in \cite{Jaramillo:2026ygy}, the existence of a maximum in $\Omega$ suppresses the magnetorotational instability (MRI) in the inner disk.
 The direct consequence is that the accretion flow stalls at $r \gtrsim r_\Omega$ owing to the inhibition of angular momentum transport, and a low-density, low-luminosity ``effective shadow'' naturally forms in the central region.
 Motivated by these GRMHD results, we adopt the following phenomenological electron number density and temperature profiles to describe the thick accretion disk surrounding the boson star.
 We treat the angular-velocity extremum radius $r_\Omega$ as the effective inner boundary of the accretion flow and use it as the reference radius for the radial profiles:
 
 \begin{equation}
 	n_e(r,\theta)
 	= n_0
 	\left(\frac{r}{r_\Omega}\right)^{-2}
 	\, \Theta(r-r_\Omega)\,
 	\exp\!\left(-\frac{z^2}{2 H^2 R^2}\right),
 \end{equation}
 
 \begin{equation}
 	T_e(r)
 	= T_0
 	\left(\frac{r}{r_\Omega}\right)^{-1},
 \end{equation}
 $n_0$ and $T_0$ are the reference electron number density and temperature at $r = r_\Omega$, respectively.
$\Theta$ denotes the Heaviside step function, ensuring that the disk is truncated inside $r_\Omega$, while $H$  is a dimensionless parameter that controls the vertical scale height of the accretion flow.
 
 The magnetic field strength is parameterized by the cold magnetization parameter $\sigma$ as
 \begin{equation}
 	B = \sqrt{\sigma\rho},\qquad \rho = n_e m_p c^2,
 \end{equation}
 where $\rho$ is the fluid mass density. In this work we set $\sigma = 0.1$.
 
 For the magnetic field configuration, a purely toroidal field is assumed:
 \begin{equation}
 	b^\mu \sim (0,\,0,\,0,\,1),
 \end{equation}
 i.e., the magnetic field winds along the $\phi$-direction and satisfies the ideal MHD condition $u_\mu F^{\mu\nu} = 0$.
 
 In this work, we consider synchrotron radiation from relativistic thermal electrons.
 For a thermal electron distribution, the synchrotron emissivity is given by
 \begin{equation}
 	j_\nu = \frac{\sqrt{3}\, n_e e^2 \nu}{6c\,\theta_e^2}\, I(x),\qquad
 	x \equiv \frac{\nu}{\nu_c},\qquad
 	\nu_c = \frac{3eB\theta_e^2\sin\theta_B}{4\pi m_e c},
 \end{equation}
 where $\theta_e = k_B T_e / (m_e c^2)$ is the dimensionless electron temperature, and $\theta_B$ is the angle between the radiation direction and the magnetic field (i.e., the pitch angle). The absorption coefficient follows from Kirchhoff's law $\alpha_\nu = j_\nu / B_\nu$, with $B_\nu$ the Planck function.
 
 In the intrinsic framework of synchrotron radiation, the emissivity depends strongly on the pitch angle $\theta_B$, making synchrotron radiation inherently anisotropic. Retaining this directional dependence amounts to using the full orientation-dependent emissivity $j_\nu(\theta_B)$, whose dimensionless function is given by the fitting formula
 \begin{equation}
 	I(x) = 2.5651 \left(1 + 1.92\, x^{-1/3} + 0.9977\, x^{-2/3}\right) \exp\!\left(-1.8899\, x^{1/3}\right).
 \end{equation}
 We adopt this anisotropic radiation scheme in the present work.

In curved spacetime, the radiation transport from the source to the observer is described by the covariant radiative transfer equation. We employ a ray-tracing method and integrate the transfer equation along null geodesics in the given spacetime metric\cite{Zhang:2024lsf,Zhang:2023bzv,Lee:2022rtg}. The evolution of the specific intensity along a ray is governed by the differential equation
\begin{equation}
	\frac{d}{d\lambda}\!\left(\frac{I_\nu}{\nu^3}\right) = \frac{j_\nu}{\nu^2} - \nu\alpha_\nu\!\left(\frac{I_\nu}{\nu^3}\right),
\end{equation}
where $\lambda$ is the affine parameter along the geodesic. The redshift factor $g \equiv \nu_0/\nu$ relates the emitted frequency $\nu$ to the observed frequency $\nu_0$, and is given by the inner product of the fluid four-velocity $u^\zeta$ and the photon four-momentum $p^\zeta$ as $g = -1/(p^\zeta u_\zeta)$. The emission and absorption coefficients $j_\nu$ and $\alpha_\nu$ are evaluated locally using the synchrotron formulas presented above, together with the local electron number density, temperature, and magnetic field strength.

\subsection{Visibility Amplitude}
\label{sec31}

This part describes the theoretical method for calculating interferometric visibility amplitudes from simulated images of boson stars. This formalism serves as the basis for the visibility analysis and model intercomparison carried out in the following section.
The visibility function is defined as the two dimensional Fourier transform of the source-plane brightness distribution $I(\mathbf{x})$:
\begin{equation}
	\mathcal{V}(\mathbf{u}) = \int I(\mathbf{x})\, e^{-2\pi i\,\mathbf{u}\cdot\mathbf{x}}\, d^2\mathbf{x},
\end{equation}
where $\mathbf{u} = (u, v)$ is the baseline vector in units of the observing wavelength.
The visibility amplitude $|\mathcal{V}(\mathbf{u})|$ and phase $\arg\mathcal{V}(\mathbf{u})$ constitute the fundamental observables of interferometry.
In very-long-baseline interferometry (VLBI) observations, the complex visibility is obtained by cross-correlating the signals from pairs of stations on each baseline, and the visibility amplitude directly encodes the Fourier power spectrum of the source brightness distribution. We therefore focus exclusively on the visibility amplitude.

In the context of gravitationally lensed sources, $I(\mathbf{x})$ is the specific intensity distribution on the observer plane obtained via ray tracing, with $\mathbf{x} = (\alpha, \beta)$ the celestial coordinates and $\mathbf{u} = (u, v)$ the baseline coordinates.
In practice, the image is represented as a two dimensional pixel array, so the continuous Fourier transform is replaced by the discrete Fourier transform (DFT):
\begin{equation}
	\mathcal{V}(u, v) = \sum_{j=1}^{N_x}\sum_{k=1}^{N_y} I(\alpha_j, \beta_k)\, e^{-2\pi i (u\alpha_j + v\beta_k)}\, \Delta\alpha\,\Delta\beta,
\end{equation}
where $N_x$ and $N_y$ are the numbers of pixels in the two coordinate directions, and $\Delta\alpha$, $\Delta\beta$ are the pixel sizes.
By evaluating $\mathcal{V}(u, v)$ along different baseline directions, one obtains the complex visibility and subsequently extracts its amplitude $|\mathcal{V}(u, v)|$.

\section{Geodesic and Lensing Structures}\label{sec3}
In this section, we investigate the null geodesic structure of boson star spacetimes and its implications for gravitational lensing. Owing to the spherical symmetry of the spacetime, the geodesic motion can be restricted to the equatorial plane $\theta=\pi/2$ without loss of generality. Under the metric ansatz introduced previously, null geodesics in the spacetime satisfy the normalization condition, which reads
\begin{equation}
	g_{\mu\nu}p^\mu p^\nu
	=
	-A(r)\dot{t}^2
	+
	B(r)^{-1}\dot{r}^2
	+
	r^2\dot{\phi}^2
	=
	0,
\end{equation}
where $p^\mu$ is the four-momentum of the photon. Due to the existence of Killing vectors associated with stationarity and axial symmetry, two conserved quantities can be introduced,
\begin{equation}
	E = A(r)\,\dot{t},
	\qquad
	L = r^2\,\dot{\phi},
\end{equation}
which correspond to the energy and angular momentum of the photon, respectively.

Substituting the conserved quantities into the null condition, it gives
\begin{equation}
	\dot{r}^2
	= E^2B(r)
	\left[
	\frac{1}{A(r)} - \frac{b^2}{r^2}
	\right],
\end{equation}
where
\begin{equation}
	b \equiv \frac{L}{E}
	\label{bd}
\end{equation}
is the impact parameter. For convenience, we define the effective potential as
\begin{equation}
	V_{\rm eff}(r) \equiv \frac{A(r)}{r^2},
	\label{eq:Veff}
\end{equation}
which determines the turning point structure and circular photon orbits of null geodesics.
Since $\dot{r}^2 \ge 0$, the allowed region of photon motion satisfies
\begin{equation}
	\frac{1}{b^2} \ge V_{\rm eff}(r).
	\label{allow}
\end{equation}
The equality in Eq.~\eqref{allow} determines the turning point $V_{\rm eff}(r_t) = \frac{1}{b^2}$, where $r_t$ denotes the radial position of the closest approach. In asymptotically flat spacetimes, photons coming from spatial infinity will first reach the turning point and subsequently are  scattered back to infinity.

Circular photon orbits correspond to degenerate turning points, where the radial velocity and acceleration vanish simultaneously. This condition requires
\begin{equation}
	V'_{\rm eff}(r) = 0.
	\label{eq:photon sphere}
\end{equation}
The set of orbits satisfying these conditions is referred to as the photon sphere. 
For a black hole, the photon sphere determines the shadow boundary and corresponds 
to the photon ring in the observed image. Moreover, the photon ring consists of a 
nested sequence of subrings whose widths decay exponentially, with the deviation 
from the critical impact parameter obeying $\Delta b_m \propto e^{-\pi m}$, where $m$ labels the half-orbit number 
(for a Schwarzschild black hole)\cite{Gralla:2019xty}.

\begin{figure}[!htbp]
	\centering
	\includegraphics[width=0.75\textwidth]{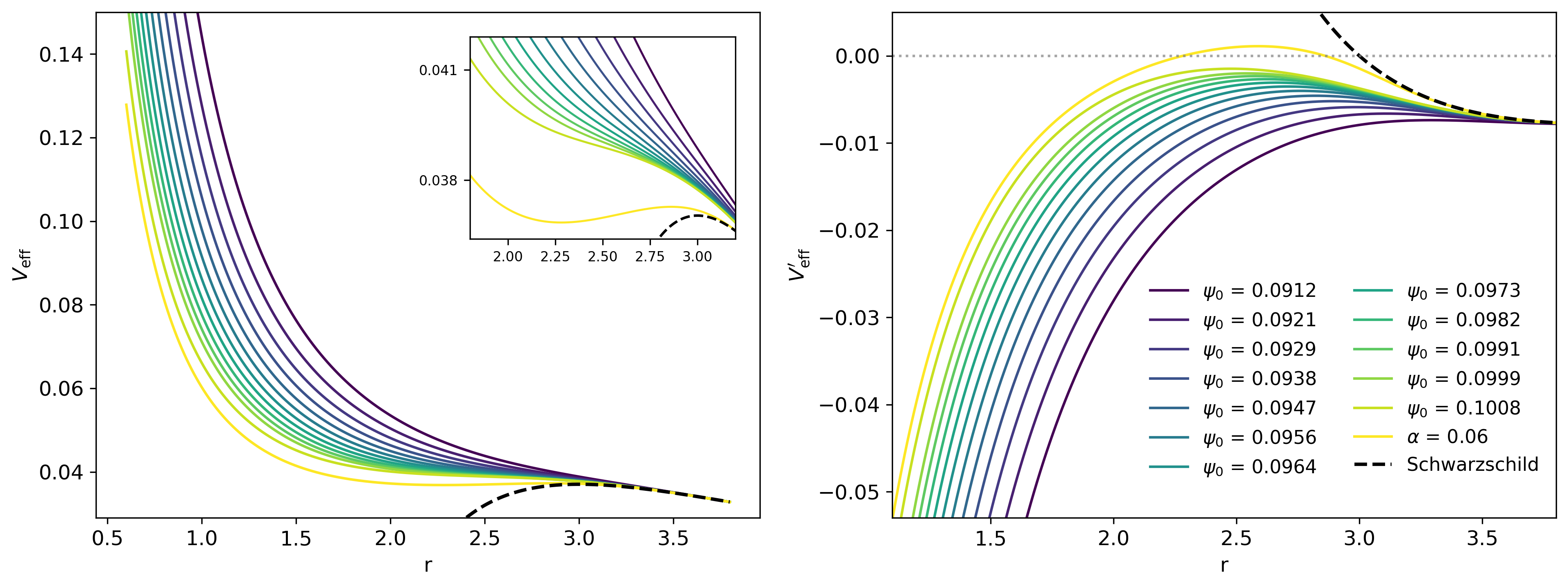}
	\caption{
		Left panel: Effective potential $V_{\rm eff}(r)$ as a function of $r$ for different boson star configurations. 
		Right panel: Radial derivative $V'_{\rm eff}(r)$ as a function of $r$. 
		The corresponding boson star configurations are indicated in the legend. 
		Curves labeled only by $\psi_0$ correspond to solutions with $\alpha=0.08$, while the unlabeled $\alpha$ curve corresponds to the configuration with $\alpha=0.06$ and $\psi_0=0.0708$, which possesses a photon ring. The black dashed curve is for the Schwarzchild black hole.
	}
	\label{eff}
\end{figure}

The left panel of Fig.~\ref{eff} shows the radial behavior of $V_{\rm eff}(r)$ for different boson star configurations, while the right panel presents its radial derivative. The colored curves correspond to boson star solutions with $\alpha=0.08$ and increasing compactness, obtained for $\psi_0$ values starting from $0.0904$, specifically, $\psi_0 = 0.0912 + 0.0008\,k$, $k=0,\dots,12$. Among these, the four configurations with $k = 0, 4, 8, 12$ are, in order of increasing compactness, those previously labeled as BS1, BS2, BS3, and BS4 in Tab.~\ref{para}. The figure also shows the configuration with $\alpha=0.06$ and $\psi_0=0.0708$, which was earlier denoted BS5 and which develops a photon ring. The black dashed curves represent the Schwarzschild spacetime, shown for comparison.

From the left panel, one can observe that $V_{\rm eff}$ grows rapidly as $r\to0$. According to Eq.~\eqref{allow}, this implies that only photons with sufficiently small $b$ can penetrate into the central region of the boson star. In combination with Definition~\eqref{bd}, this means that photons traversing the interior region tend to move along nearly radial trajectories with very small angular momentum. Consequently, they scarcely interact with the accretion disk, giving rise to a shadow structure in the central region of the image.

In Fig.~\ref{eff}, $V_{\rm eff}$ with $\alpha=0.08$ decreases monotonically with $r$, therefore, no circular photon orbit exists. As $\psi_0$ increases, however, the configurations 
become more compact and $V_{\rm eff}$ gradually develops an increasingly flat region. 
This trend is also visible in the right panel: as $\psi_0$ increases, $V'_{\rm eff}$ 
forms a peak, towards zero.
Due to the presence of this flat region in the effective potential, photons with 
suitable impact parameters maintain an extremely small $\dot{r}$ throughout the entire 
flat region. Their radial motion is strongly suppressed, allowing them to spend a 
large amount of affine time there; meanwhile, the angular motion continues, enabling 
the photons to undergo large-angle deflections and form lensing rings in the image. 
As the compactness increases, the radial extent of the flat region grows, and the 
affine time spent within it becomes even longer. This allows photons with smaller 
angular momenta to accumulate sufficient angular deflection in the same region, 
leading to a broadening of the lensing rings. Therefore, unlike the lensing rings of black holes, which are highly concentrated near the critical impact parameter, the lensing rings in boson star images are expected to exhibit a much broader spatial distribution.

For the particularly compact configuration with BS5 in Tab.~\ref{para}, the effective potential develops both a local maximum and a local minimum. Correspondingly, in the $V'_{\rm eff}$ profile, the curve, which is highlighted in yellow, crosses zero twice. The local maximum corresponds to an unstable photon ring, analogous to the Schwarzschild photon sphere, with the crucial difference that photons crossing it in a black hole are captured by the horizon, whereas in a horizonless boson star they can reach a turning point and travel back. More interestingly, the local minimum corresponds to a stable circular photon orbit. Photons emitted near this region may be confined within the effective potential well for a certain range of impact parameters. This behavior has no counterpart in the Schwarzschild black hole.

It is worth noting that photons forming higher-order lensing rings spend a longer affine time propagating through the strong-field region. Consequently, even a minute variation in the impact parameter \(b\) can accumulate into a significant change in the total deflection angle, leading to progressively narrower high-order rings in the image plane. 
Furthermore, as the compactness of the boson star increases, the peak of $V'_{\rm eff}$ gradually approaches zero and eventually reaches zero. In this regime, the radial motion of photons becomes increasingly dominated by the second derivative of \(\dot{r}\), and the dynamics begin to resemble those near a black-hole photon sphere. Accordingly, the narrowing of the highest-order lensing rings approaches an exponential scaling.

\begin{figure}[!htbp]
	\centering
	\includegraphics[width=0.50\textwidth]{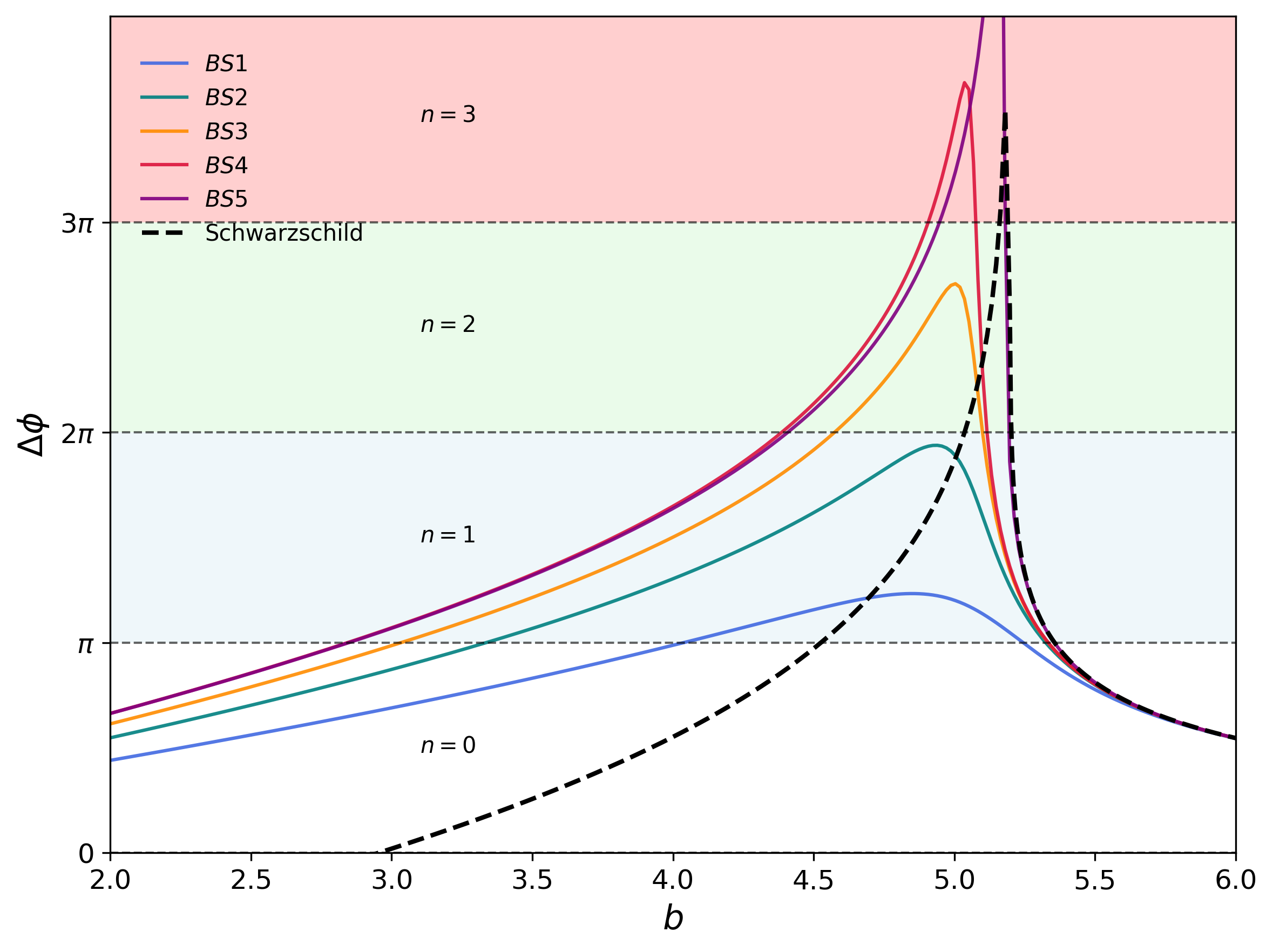}
	\caption{
		Deflection angle $\Delta\phi$ as a function of the impact parameter $b$ for five boson star configurations with increasing compactness listed in Tab.~\ref{para}. Horizontal dashed lines denote $\Delta\phi=\pi n$ with $n=1,2,3$, corresponding to photon trajectories completing one, two, and three half-orbits around the compact object before escaping to infinity.
	}
	\label{dphi}
\end{figure}

To demonstrate clearly that boson stars possess wider lensing rings, We compute the deflection angle of null geodesics of different configuration, which can be written as
\begin{equation}
	\Delta\phi(b)
	=	2\int_{r_t}^{\infty}\frac{\dot{\phi}}{\dot{r}}dr-\pi=
	2\int_{r_t}^{\infty}
	\frac{dr}{
		r^2\sqrt{
			B(r)\left(
			\dfrac{1}{A(r)b^2}
			-
			\dfrac{1}{r^2}
			\right)
	}}
	-\pi ,
\end{equation}
where \(r_t\) is the turning point of the photon trajectory.

Fig.~\ref{dphi} shows the deflection angle \(\Delta\phi(b)\) for five different boson star configurations as listed in Tab.~\ref{para} and Schwarzschild spacetime. Following the conventional lensing classification, we define \(n=1\) as photon trajectories completing one half-orbit around the compact object, namely
\begin{equation}
	\Delta\phi \ge \pi n .
\end{equation}
As the compactness of the boson star increases, the peak structure of the deflection angle $\Delta\phi(b)$ becomes progressively sharper, and the corresponding impact parameter intervals for $n=1,2,3$ broaden significantly. As a result, strong lensing orbits are no longer confined to a narrow range of impact parameters but instead occupy an extended region, ultimately producing broader and more diffuse lensing rings in the image. Since these photons traverse longer paths within the radiating material, the imaging structure of boson stars is more strongly dependent on the distribution of the emission source than in the black hole case.

It is also worth noting that the nomenclature for photon rings adopted in this paper differs from the conventional definition. In the conventional definition, $n=0$ corresponds to the direct image, $n=1$ is called the lensing ring, and $n\geq2$ are referred to as photon rings. However, most of the boson stars we study do not possess a photon sphere necessary for the formation of photon rings. Nevertheless, due to the aforementioned lensing effect, they still produce a large number of photons that undergo multiple windings. Therefore, in this paper we uniformly refer to all emission rings with $n\geq1$ as lensing rings.
\section{Numerical Result}\label{sec4}
In this section, we perform numerical simulations of imaging and visibility amplitudes for compact boson stars, employing the methods described in the preceding sections. Our earlier analysis of their lensing structure demonstrated that boson stars form substantially broader $n>1$ rings, which are expected to be highly sensitive to the distribution of the radiating source. To demonstrate the sensitivity of the imaging results to the distribution of the 
accretion flow, we perform simulations with two disk thickness values, $H=0.1$ and 
$H=1$, by the approach adopted in \cite{Vincent:2022fwj}. For the imaging, we adopt a resolution of $512 \times 512$ pixels.
\subsection{Thick Disk Image}
\label{sec41}

\begin{figure}[!htbp]
	\centering
	\includegraphics[width=0.99\textwidth]{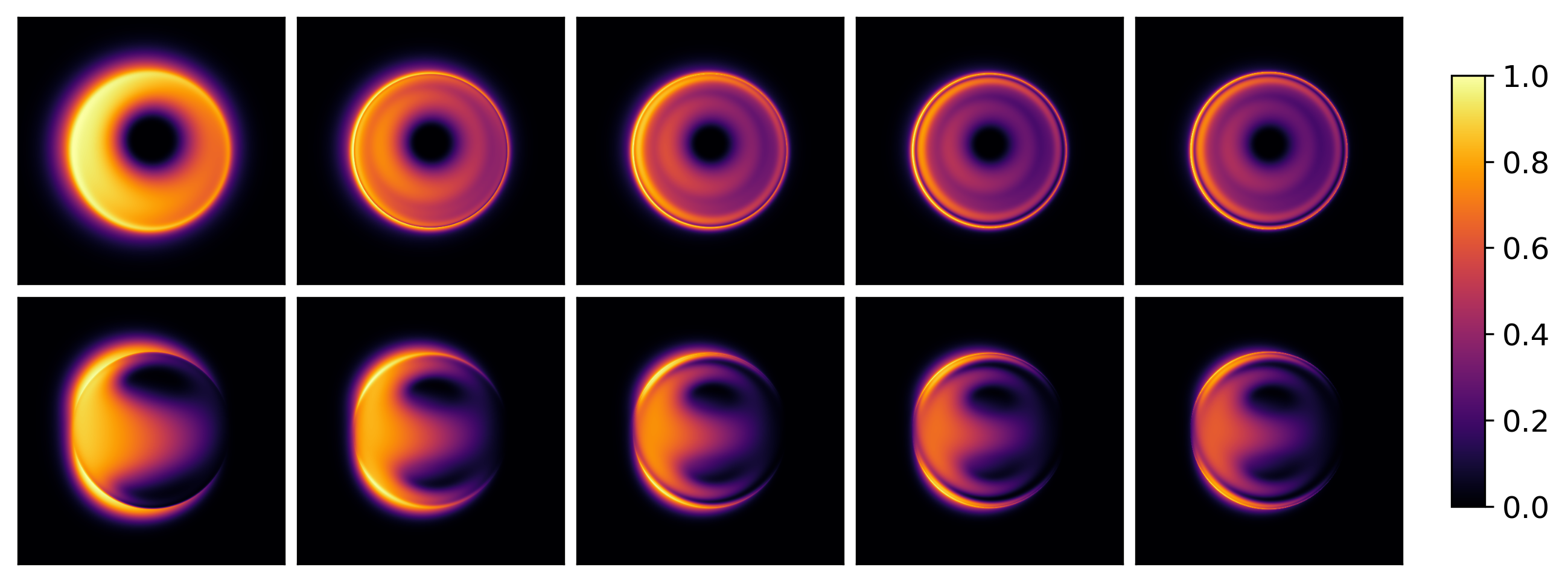}
	\caption{
		Images of the thick disk with $H = 1$. 
		The top row corresponds to an inclination angle of $17^\circ$, while the bottom row of $80^\circ$. 
		From the left to right, the columns show the images through BS1 to BS5.	}
	\label{2i}
\end{figure}

Fig.~\ref{2i} presents the imaging results for $H=1$. The upper and lower panels correspond to the  observer inclination angles $\theta_o= 17^\circ$ and $80^\circ$, respectively, while the columns from left to right represent the BS1–BS5 configurations with progressively increasing compactness. To better highlight the brightness distribution features, the intensity is normalized according to
$
I_{\rm norm}=\frac{I-I_{\min}}{I_{\max}-I_{\min}},
$
and all images shown correspond to the normalized intensity distribution $I_{\rm norm}$.

We first focus on the low inclination case. BS1 exhibits a relatively uniform and overall bright intensity distribution. This behavior originates mainly from its relatively loose spacetime structure, in which only the $n=1$ lensing ring is present. The path lengths traversed by different photons through the accretion flow are relatively similar, resulting in nearly uniform radiative accumulation and consequently the absence of pronounced brightness stratification. As the compactness increases, photons with large deflection angles begin to appear. In the main body of the image, the brightest region gradually concentrates toward an outer ring-like structure, which is the $n=2$ lensing ring. Meanwhile, higher-order lensing rings emerge as distinct features that separate from the main emission region, and both their degree of separation and their fractional contribution to the total intensity increase with compactness. For the high-inclination case with $\theta_{\rm o}=80^\circ$, the strong gravitational lensing effect makes two shadows observable in the image. In addition, a noteworthy phenomenon is that the $n=2$ lensing ring separates from the main radiation region. The bright ring at the outer part of the image exhibits a layered structure, and its inner portion is the $n=2$ lensing ring.
\begin{figure}[!htbp]
	\centering
	\includegraphics[width=0.99\textwidth]{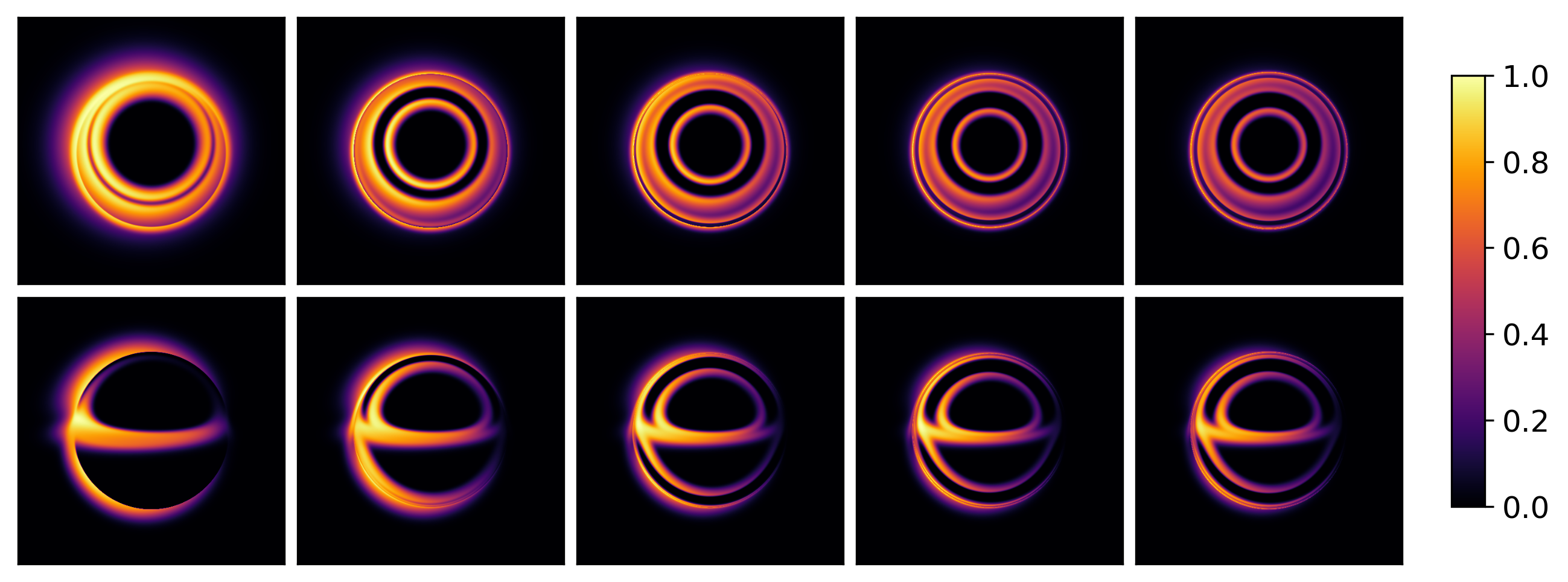}
	\caption{
	Images of the thick disk with $H = 0.1$. 
	The top row corresponds to an inclination angle of $17^\circ$, while the bottom row of $80^\circ$. 
	From the left to right, the columns show the images through BS1 to BS5.
	}
	\label{3i}
\end{figure}

Fig.~\ref{3i} shows the results for $H=0.1$. Compared with the $H=1$ case, the image exhibits several notable differences. First, the $n=2$ photon ring no longer displays a prominent brightness enhancement. Second, at low inclination, the central bright ring splits into two distinct structures: an inner narrow ring and an outer broad ring, resulting in a significant change in the image topology. At high inclination, a similar splitting feature also appears in the upper part of the image.

These two bright rings are both produced by the $n=1$ lensing ring and correspond to photons with deflection angles greater than $\pi$, meaning that they cross the equatorial plane twice during their propagation. The associated photon trajectories can be divided into three categories. Some photons pass through the high emissivity region during their first equatorial crossing, giving rise to the outer broad ring. Others pass through the high emissivity region during their second equatorial crossing, forming the inner narrow ring. In addition, there exists a population of photons whose entire turning process takes place within the low emissivity region inside the boson star and therefore contributes negligibly to the observed radiation. The combined effect of these three classes of trajectories produces a dark gap with nearly zero intensity between the two bright rings, ultimately giving rise to the clearly resolved double ring structure.

\begin{figure}[!htbp]
	\centering
	\includegraphics[width=0.65\textwidth]{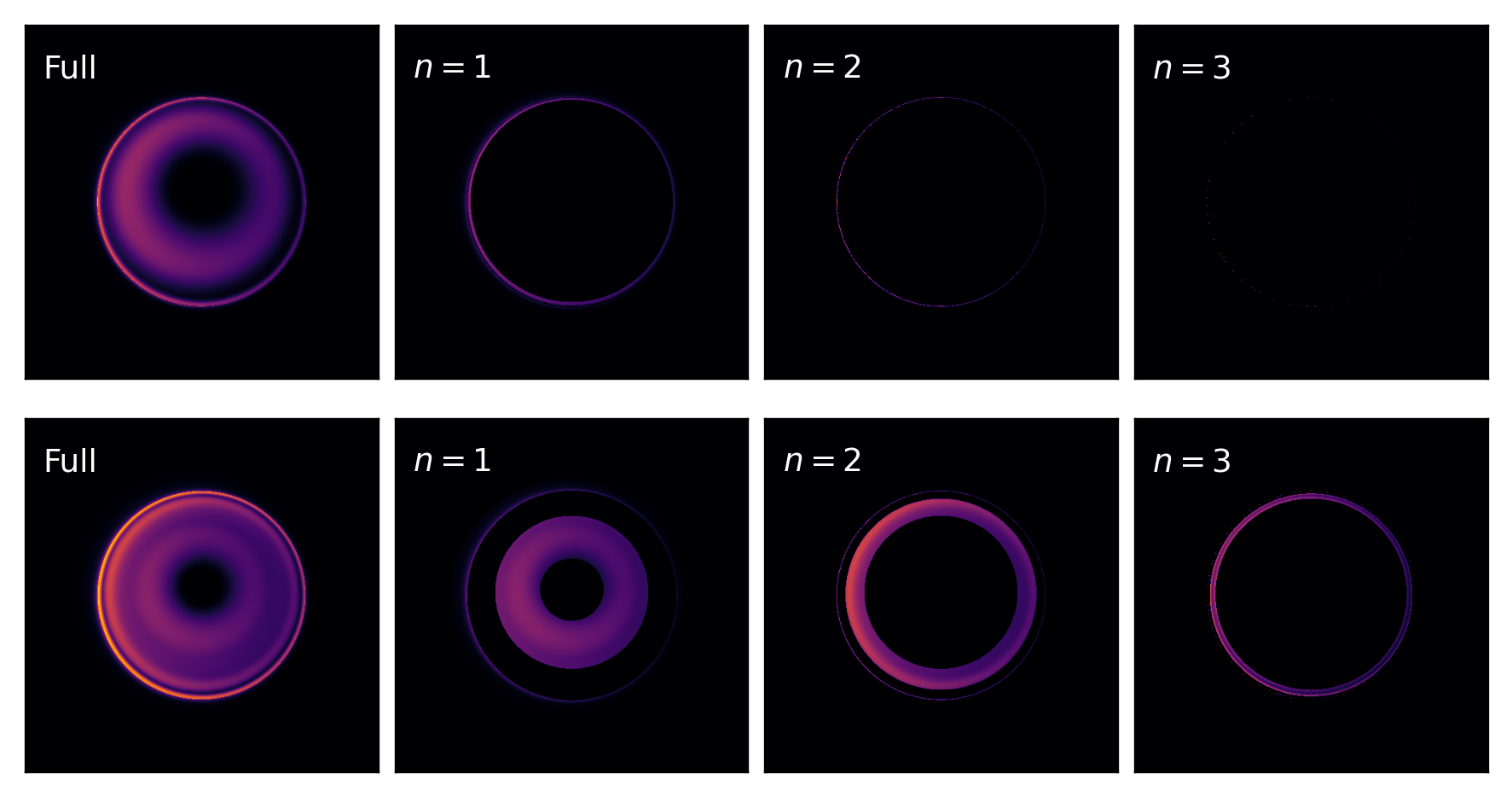}
	\caption{
		Decomposed images of the Schwarzschild black hole and BS5 for the overall intensity and the $n=1,2,3$ lensing rings. The upper row corresponds to the Schwarzschild black hole case, while the lower row corresponds to BS5.
	}
	\label{4i}
\end{figure}

To more intuitively reveal the lensing structure of the boson star, we compare in Fig.~\ref{4i} the decomposed images of a Schwarzschild black hole and BS5 for $H=1$. The upper row corresponds to the Schwarzschild results: the $n=1$ ring already converges to an extremely narrow shape, and the $n=2$ ring is even narrower; due to the limited image resolution, the $n=3$ image can no longer form a complete ring structure. The lower row for BS5 behaves differently: its $n=1$ ring is much broader than in the Schwarzschild case, and the overall intensity distribution of the ring differs significantly; even at $n=3$, the ring still maintains a relatively broad and non-uniform distribution.

\begin{figure}[H]
	\centering
	\includegraphics[width=0.95\textwidth]{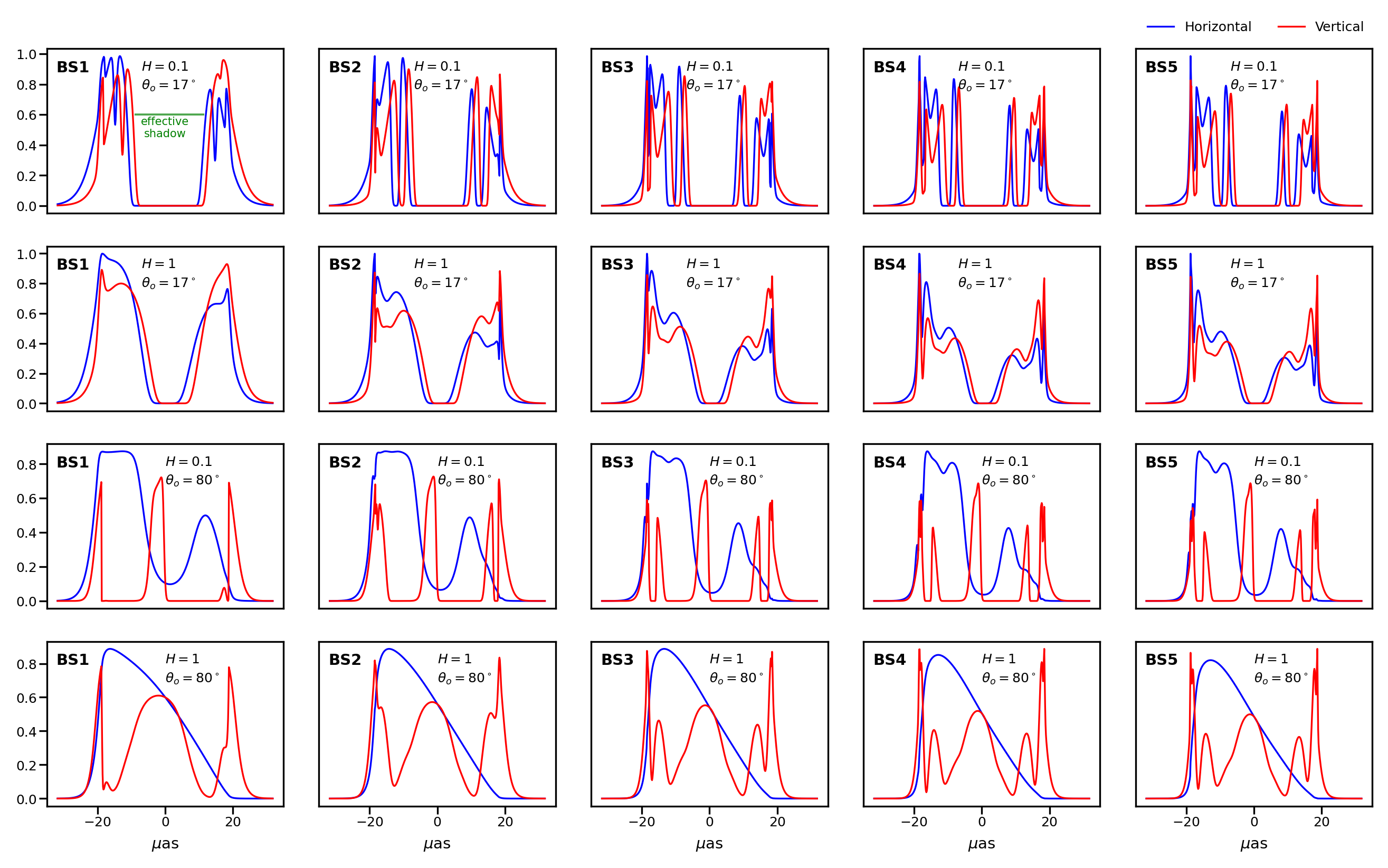}
	\caption{Intensity distributions along the $x$-axis (blue curve) and $y$-axis (red curve). The corresponding boson star model, disk thickness, and observation inclination angle are indicated in the figure.}
	\label{xyi}
	\end{figure}

Fig.~\ref{xyi} displays the intensity profiles along the $x$ and $y$ axes, with the corresponding parameters indicated in the figure. 
First, for the $\theta = 17^\circ$ case, the central dip, which corresponds to the effective shadow region becomes visibly narrower as the disk thickness increases, indicating a reduction in the shadow size. This behavior arises from the radiative contribution of the off-equatorial accretion flow. 
Furthermore, in the thin disk case with $H=0.1$, as the compactness decreases, a region where the intensity approaches zero appears near the shadow. This feature exactly matches the ring-separation phenomenon described earlier. 
For the $\theta = 80^\circ$ inclination, the intensity distribution along the vertical direction exhibits two dips. This is because the adopted accretion flow is mostly concentrated close to the boson star, and the strong gravitational lensing allows us to see the image of the lower side, thereby revealing the lower shadow as well. 
Finally, it is worth noting that in the bottom row corresponding to $H=0.1$ and $\theta=80^\circ$, a careful inspection of the peak positions reveals that the originally single intensity peak gradually splits into two distinct peaks as the compactness increases. This feature has a clear counterpart in the corresponding images. For sufficiently compact configurations, the $n=2$ lensing ring gradually separates from the higher order lensing rings. The underlying mechanism is identical to that responsible for the splitting of the $n=1$ lensing ring discussed above, these photons only traverse the low-emissivity region inside the boson star.

\subsection{Visibility}
In this section, we present the results of the visibility calculation. 
The zero baseline visibility of the accretion disk image is given by
\begin{equation}
	\mathcal{V}(0,0) = \sum_{j=1}^{N_x}\sum_{k=1}^{N_y} 
	I(\alpha_j,\beta_k)\, \Delta\alpha\,\Delta\beta,
\end{equation}
which equals the total intensity of the image. The image is first normalized 
by this total intensity, so that the normalized zero-baseline visibility 
amplitude is set to unity. The normalized image is then transformed to the 
Fourier domain via a discrete Fourier transform, yielding the two-dimensional 
complex visibility distribution $\mathcal{V}(u,v)$. The Fourier transform is performed using the Fast Fourier Transform implementation provided by NumPy \cite{numpy}.
We define the baseline length as $\mathbf{b} = \sqrt{u^2 + v^2}$,
and azimuthally average the visibility amplitude
$|\mathcal{V}(u,v)|$ at each baseline length, thereby
obtaining the azimuthally averaged visibility amplitude
as a function of $\mathbf{b}$, denoted $|\bar{\mathcal{V}}(\mathbf{b})|$.

\begin{figure}[!htbp]
	\centering
	\includegraphics[width=0.90\textwidth]{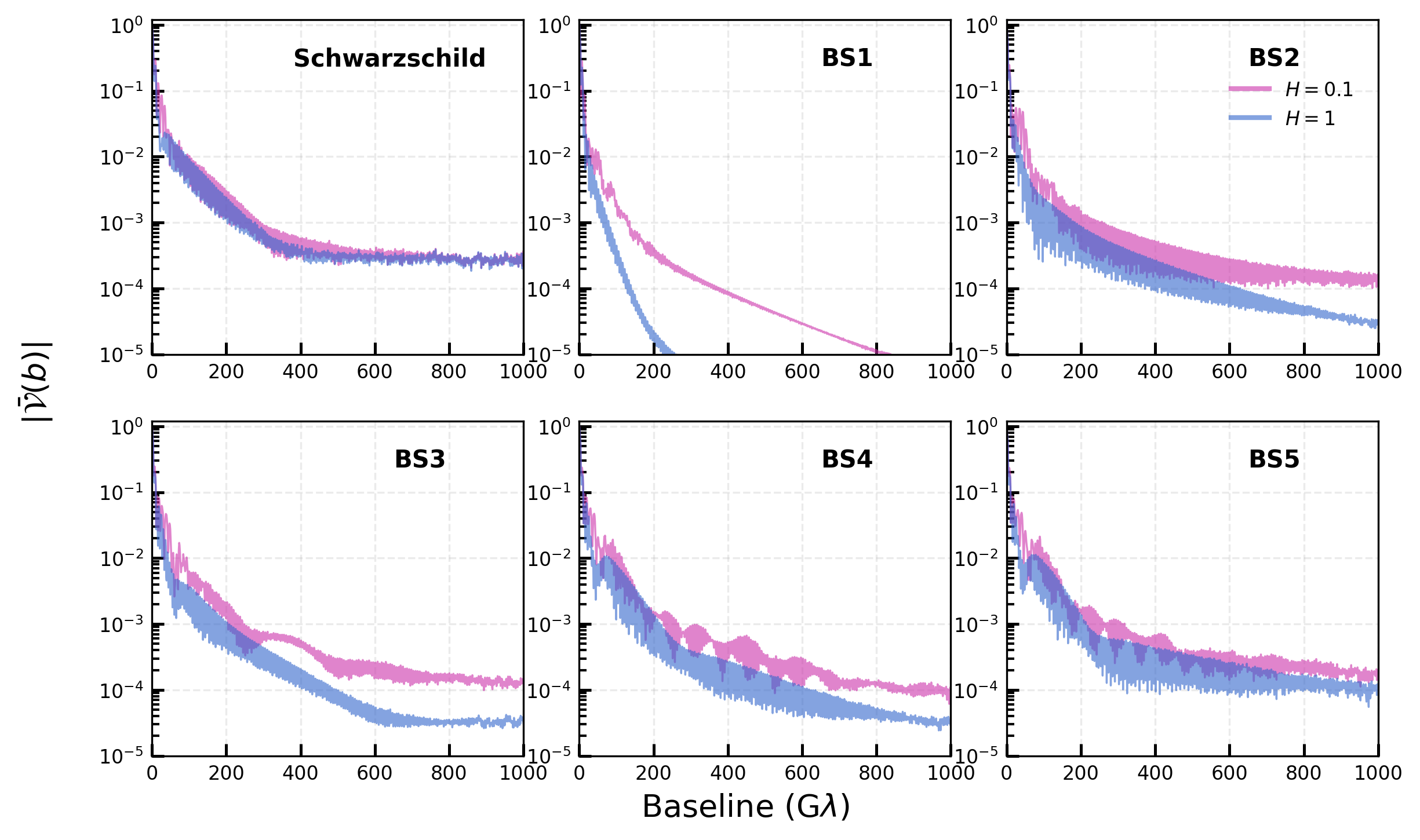}
	\caption{
		$\bar{\mathcal{V}}(\mathbf{b})$ as functions of baseline length $ \mathbf{b}$ for the Schwarzschild and BS1-5 viewed at $\theta_o=17^\circ$. In each panel, the magenta and blue curves correspond to $H=0.1$ and $H=1$ configurations, respectively. }
	
	\label{v17}
\end{figure}
\begin{figure}[!htbp]
	\centering
	\includegraphics[width=0.90\textwidth]{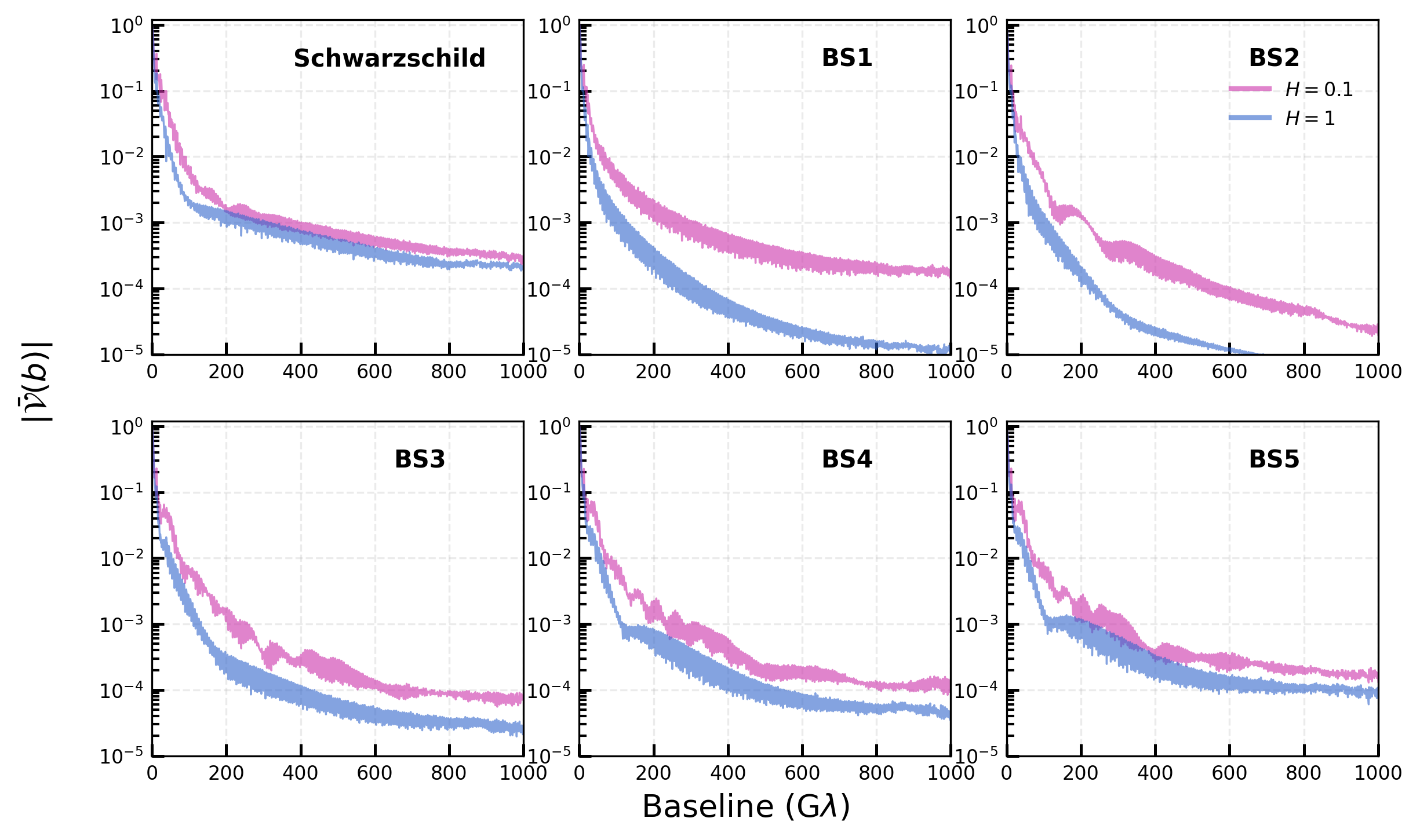}
	\caption{
	$|\bar{\mathcal{V}}(\mathbf{b})|$ as functions of baseline length $ \mathbf{b}$ for the Schwarzschild and BS1-5 viewed at $\theta_o=80^\circ$. In each panel, the magenta and blue curves correspond to $H=0.1$ and $H=1$ configurations, respectively.
	}
	\label{v80}
\end{figure}

In Fig.~\ref{v17} and Fig.~\ref{v80}, we show $|\bar{\mathcal{V}}(\mathbf{b})|$ 
for inclinations $\theta_{\rm o} = 17^\circ$ and $\theta_{\rm o} = 80^\circ$, respectively. 
At low inclination, a clear trend emerges. In the Schwarzschild background, 
$|\bar{\mathcal{V}}(\mathbf{b})|$ is only weakly sensitive to the disk thickness. 
By contrast, for boson stars the $|\bar{\mathcal{V}}(\mathbf{b})|$ curves corresponding to different 
disk thicknesses exhibit a pronounced separation, which gradually diminishes 
with increasing compactness. As noted in \cite{Johnson:2019ljv}, the long-baseline 
visibility of a Schwarzschild black hole is dominated by the exponentially 
decaying photon rings. The morphology and brightness of the photon ring are 
dictated primarily by the spacetime geometry rather than by the details of the 
accretion flow. The disk thickness does, however, 
influence the brightness distribution in the image region dominated by the 
lensing ring and the direct image. A thicker disk provides additional 
off equatorial emission that shrinks the effective size of the central shadow. 
Consequently, for the Schwarzschild case, a clear separation between the 
$H=1$ and $H=0.1$ curves appears at very short baselines.

For BS1, $|\bar{\mathcal{V}}(\mathbf{b})|$ for the two disk thicknesses show a pronounced separation. This mainly stems from the fact that the BS1 configuration is relatively diffuse and the image is dominated by the $n=1$ lensing ring. In the $H=0.1$ case, due to the absence of off equatorial emission, the central dark region becomes larger and the brightness transition at its boundary becomes sharper, thereby increasing the overall visibility amplitude. As the compactness increases, higher order lensing rings begin to emerge and gradually separate from the main body of the image. Consequently, increasingly fine structures develop in the image, leading to an enhancement of the visibility amplitude at long baselines. Moreover, increasing compactness generally raises the relative brightness of these higher order rings, further boosting the visibility amplitude at long baselines. However, this brightness enhancement mechanism is weaker in the $H=1$ case, so the degree of high frequency enhancement differs between the thin disk and thick disk models, causing their visibility curves to gradually approach each other.

It is worth noting that the image results presented in Sec.~\ref{sec41} show that, as the compactness increases, the $n=1$ lensing ring in the thin disk case splits into two distinct components. The superposition of the Fourier components associated with these two rings of slightly different radii gives rise to a beating effect. The small difference in their oscillation frequencies produces a low-frequency envelope that modulates the visibility amplitude. In the visibility amplitude profile, this appears as alternating enhancement and suppression of the oscillation, resembling a standing wave pattern. The visibility beating phenomenon has been discussed in detail in \cite{Wang:2025hzu}, where it was studied in the context of double photon ring images. Although the structures considered here correspond to a split $n=1$ lensing ring rather than two distinct photon rings, the underlying mechanism is identical, namely the superposition of visibility amplitude components associated with two ring-like structures of slightly different radii. This effect is particularly pronounced in the BS4 and BS5 images. As a result, although the $|\bar{\mathcal{V}}(\mathbf{b})|$ curves for $H=1$ and $H=0.1$ are relatively close overall, they still exhibit noticeable differences due to the beating pattern induced by the double ring structure.

At high inclination, the $|\bar{\mathcal{V}}(\mathbf{b})|$ curves display a similar overall trend to those at low inclination, but with a systematic offset between the $H=1$ and $H=0.1$ curves: the visibility amplitude of the thick disk is consistently lower than that of the thin disk. We attribute this offset to the enhanced off-equatorial radiation of the thick disk at high inclination, which contributes significantly more total flux. Consequently, in the normalized $|\bar{\mathcal{V}}(\mathbf{b})|$, the thick disk has a larger normalizing denominator, which depresses the entire curve. 

Finally, we observed the presence of additional oscillatory features in the 800--1000~G$\lambda$ range. We attribute these features to numerical artifacts associated with the finite image resolution. As discussed in Sec.~\ref{sec3}, the widths of higher-order lensing rings decrease with increasing ring order for both black holes and boson stars. Once the ring width falls below the pixel scale, the ring structure can no longer be properly resolved and is represented by only a handful of pixels. Such undersampling introduces aliasing and other discretization effects in the discrete Fourier transform, leading to spurious high-frequency oscillations in the visibility amplitudes.
\section{Conclusions and Discussions}\label{sec5}

In this work, we systematically investigated the geodesic structure, gravitational lensing properties, accretion flow images, and visibility amplitudes of compact boson stars. The key result of this work was that the flat effective potential of compact boson stars induced a systematic broadening of lensing rings, which in turn enhanced the imprint of accretion flow structure in both imaging and interferometric observables. By selecting boson star configurations with different degrees of compactness, we analyzed how the spacetime geometry shaped the image morphology and the corresponding
interferometric observables.

At the geodesic level, we found that the effective potential of compact boson stars possessed a rather flat region. This flat region significantly suppressed the radial motion of photons, allowing a wider range of impact parameters to accumulate sufficient azimuthal deflection. As the compactness of the boson star increased, the radial extent of this flat region broadened, causing lensing rings of all orders to become systematically wider. When the compactness became sufficiently high, the effective potential even developed both stable and unstable circular photon orbits. Notably, these features were not unique to boson stars; for some horizonless compact objects, similar behavior also emerged as their compactness increases.

Based on the geodesic properties discussed above, we further performed ray tracing simulations for accretion flows with different disk thicknesses. The results revealed markedly different imaging features for different disk geometries. In the thin disk case ($H=0.1$), the central bright ring split into two clearly separated subrings, with a pronounced dark gap between them, resulting in a significant change in the image topology. This feature arose from the substantial broadening of the $n=1$ lensing ring. For a subset of the associated photon trajectories, the turning points lay within the low emissivity region inside the boson star, where only negligible radiation was accumulated along the trajectory. Consequently, a brightness deficit developed within the lensing ring, leading to the emergence of a double ring
structure.

In the Fourier domain, we computed the visibility amplitudes and performed a
systematic comparison with the Schwarzschild black hole case. The results
showed that the visibility amplitudes of boson stars exhibited a much
stronger sensitivity to the accretion flow structure than those of black
holes, as evidenced by the pronounced separation between the visibility
curves corresponding to different disk thicknesses. As the compactness
increased, higher order lensing rings contributed an increasingly larger
fraction of the total radiation, thereby reducing the influence of the
accretion flow geometry on the image structure and causing the visibility
curves for different disk thicknesses to gradually converge. However, in the
thin disk case, sufficiently compact configurations caused the $n=1$ lensing
ring to split into two distinct subrings. The resulting double ring structure
produced significant interference in the Fourier domain, which manifested
itself as a pronounced beating modulation in the visibility amplitude.
Consequently, although the overall visibility amplitudes of different models
became increasingly similar with increasing compactness, their oscillatory
patterns remained significantly different due to the beating effect induced
by the double ring structure.

In summary, our results demonstrated that the rich and complex lensing structures of compact boson stars could give rise to observable signatures that differed significantly from those of black holes. Owing to their stronger sensitivity to the spatial distribution of radiation in the accretion flow, these lensing features encode abundant information about the underlying spacetime geometry in both the image and visibility domains. Future high resolution very long baseline interferometric observations may exploit these signatures to place more stringent observational constraints on horizonless compact object models such as boson stars, and provide a new avenue for assessing their viability as alternatives to black holes.

\section*{Acknowledgments}
We are grateful to Zhenyu Zhang for discussions. This work was partially supported by the National Natural Science Fundation of China (Grant No. 12575048 and 12175008). M. Guo is also supported by the BNU Tang Scholar.

\bibliographystyle{utphys}
\bibliography{draft}

\end{document}